\documentclass[12pt]{article}
          \usepackage{amsfonts}
          \usepackage{amssymb}

\newtheorem{thm}{Theorem}
\newtheorem{cor}[thm]{Corollary}
\newtheorem{lemma}[thm]{Lemma}

\def\qed{{\bf QED}}

\def\tr{\hbox{Tr}}

\def\be{\begin{eqnarray}}
\def\ee{\end{eqnarray}}
\def\bee{\begin{eqnarray*}}
\def\eee{\end{eqnarray*}}
\def\bmx{\begin{pmatrix}}
\def\emx{\end{pmatrix}}

\def\ts{\textstyle}

\def\rt2{\ts \frac{1}{\sqrt{2}} }

\def\ot{\otimes}

\title{Hypercontractivity for semigroups of unital qubit channels}

    \author{Christopher King\\
    \\
{\small      Department of Mathematics} \\
{\small      Northeastern University} \\
{\small      Boston MA 02115}}

\begin{document}

     \maketitle

\begin{abstract}
Hypercontractivity is proved for products of qubit channels that belong to self-adjoint semigroups.
The hypercontractive bound gives necessary and sufficient conditions for a product of the form
$e^{- t_1 H_1} \ot \cdots \ot e^{ - t_n H_n}$ to be a contraction from $L^p$ to $L^q$, where $L^p$ is the algebra of
$2^n$-dimensional matrices equipped with the normalized Schatten norm, and each generator $H_j$ is a self-adjoint
positive semidefinite operator on the algebra of $2$-dimensional matrices. As a particular case the result establishes the hypercontractive bound for a product of qubit depolarizing channels.
\end{abstract}

\section{Introduction}\label{intro}
Hypercontractivity concerns the contractive properties of semigroups
$e^{ - t H}$ from $L^p$ to $L^q$ with $q > p$. This notion first came to light in the
work of Nelson \cite{Nel1}, \cite{Nel2} on quantum field theory,  and was subsequently investigated by
mathematical physicists in a variety of settings \cite{Glimm}, \cite{Federbush}, \cite{Gr2}, \cite{HKS}, \cite{CL1}, \cite{Car1}. These investigations led to the logarithmic Sobolev inequalities \cite{Gr1} and other mathematical advances \cite{Be}, \cite{Bi}, as well as to applications in theoretical computer science \cite{KKL} and other fields (the article \cite{Gr3} provides a useful historical overview). Recently hypercontractivity has been applied to some problems of interest in quantum information theory (QIT), as described for example in \cite{BRW}, \cite{MO1}, \cite{KT}, \cite{M1}. In QIT the natural setting for hypercontractivity 
is quantum channel semigroups on full matrix algebras, and that is the subject of this paper. We consider the simplest non-trivial example, namely products of qubit channels, and prove that hypercontractivity holds for this case. 

\medskip
In order to state our results we first review some notation and definitions.
For each $k \ge 1$, we denote by ${\cal M}_k$ the algebra of $k \times k$ complex-valued matrices.
${\cal L}({\cal M}_k)$ will denote the space of linear
maps on ${\cal M}_k$ (sometimes called superoperators):
\be
{\cal L}({\cal M}_k) = \{L \,:\, {\cal M}_k \rightarrow {\cal M}_k \}
\ee
The algebra ${\cal M}_k$ is equipped with the inner product $\langle A, B \rangle = \tr A^* B$, and this provides the definition of the adjoint $\widehat{L}$ of a map $L \in {\cal L}({\cal M}_k)$:
\be
\langle A, L(B) \rangle = \langle \widehat{L}(A), B \rangle
\ee
We will say a map $L$ is self-adjoint if $L = \widehat{L}$, and 
positive semidefinite (denoted $L \ge 0$) if $L = \widehat{L}$ and
$\langle M, L(M) \rangle \ge 0$ for all $M \in {\cal M}_k$.

\medskip
We write $A \ge 0$ to indicate that $A \in {\cal M}_k$ is hermitian and positive semidefinite.
A map $L$ is called positivity preserving if $L(A) \ge 0$ whenever $A \ge 0$.
Recall that a map $L \in {\cal L}({\cal M}_k)$ is {\em completely
positive} \cite{Stine}  if ${\cal I}_m \ot L$ is positivity preserving on ${\cal M}_{m} \ot {\cal M}_{k}   = {\cal M}_{m k }$ for all $m \ge 1$,
where ${\cal I}_m$ denotes the identity map in ${\cal L}({\cal M}_m)$. 
Finally a {\em quantum channel} on $\mathbb{C}^k$ is a completely positive trace-preserving (CPTP) element of ${\cal L}({\cal M}_k)$.

\medskip
The main result of this paper concerns the contractive properties of quantum channels.
Contractivity is defined using norms on matrix algebras.
Recall that for $M \in {\cal M}_k$, and all $p \ge 1$, the Schatten norm of $M$ is defined to be
\be\label{def:||-norm}
|| M ||_p = \left(\tr |M|^p \right)^{1/p}
\ee
where $\tr$ is the standard trace on ${\cal M}_k$.
We will also use the normalized version of the trace on ${\cal M}_k$, which we denote by $\tau$, and define by
\be
\tau(M) = \frac{1}{k} \, \tr M
\ee
$\tau$ gives rise to normalized versions of the Schatten norm, which we denote by $||| M |||_p$ and define as
\be\label{def:|||-norm}
||| M |||_p =  \left(\tau \left(|M|^p\right) \right)^{1/p} = k^{-1/p} \, || M ||_p
\ee
Note that $|| M ||_p$ is non-increasing as a function of $p$, while $||| M |||_p$ is non-decreasing as a function of $p$. Also
$||| I_k |||_p = 1$ where $I_k$ is the identity matrix in ${\cal M}_k$.

\medskip
For $L \in {\cal L}({\cal M}_k)$ we define families of norms using both (\ref{def:||-norm}) and (\ref{def:|||-norm}):
for all $p,q \ge 1$
\be\label{def:map-norm}
|| L ||_{p \rightarrow q} = \sup_M \frac{|| L(M) ||_q}{|| M ||_p}, \quad
||| L |||_{p \rightarrow q} = \sup_M \frac{||| L(M) |||_q}{||| M |||_p}
\ee
where in both cases the supremum is taken over all $M \in {\cal M}_k$.
Note that these norms are related by
\be\label{norms-relate}
||| L |||_{p \rightarrow q} = k^{-1/q + 1/p} \, || L ||_{p \rightarrow q}
\ee

\medskip
A {\em qubit channel} is a CPTP map in ${\cal L}({\cal M}_2)$. 
We shall be concerned with unital qubit channels which belong to semigroups with self-adjoint generators.
Accordingly we define the following set of generators:
\be\label{def:G}
G = \{H \in {\cal L}({\cal M}_2) \,:\, H(I_2)=0, \, H = \widehat{H}, \, H \ge 0 \}
\ee
where $I_2$ denotes the $2 \times 2$ identity matrix. Given $H \in G$
we write $H^n$ for the $n$-fold composition of the map $H$, and then define the semigroup $e^{ - t H}$ by
\be
e^{ - t H}(M) = \sum_{n=0}^{\infty} \frac{(-t)^n}{n!} \, H^n(M), \quad M \in {\cal M}_2
\ee
The condition that $e^{ - t H}$ be completely positive for all $t \ge 0$ puts further constraints on $H$,
so we define the set of generators of CP semigroups by
\be
G^{CP} = \{ H \in G \,:\, e^{ - t H} \,\,\mbox{is CP for all} \,\, t \ge 0 \}
\ee
We also define the `unit rate' generators
\be
G^{CP}_1 = \{ H \in G^{CP} \,:\, \| H \|_{\min} = 1 \}
\ee
where
\be
\| H \|_{\min} = \inf_{{\rm Tr} M = 0} \, \frac{\langle M, H(M) \rangle}{\langle M, M \rangle}
\ee
We can now state our main hypercontractivity bound.

\begin{thm}\label{thm:main}
Let $H_1,\dots,H_n \in G^{CP}_1$. For all $1 < p \le q$,
\be\label{main1}
||| e^{ - t_1 H_1} \ot \cdots \ot e^{- t_n H_n} |||_{p \rightarrow q} =1 \,\, \mbox{if and only if} \,\, \max_i e^{ - t_i} \le \sqrt{\frac{p-1}{q-1}}
\ee
\end{thm}

\medskip
One channel of particular interest is the qubit depolarizing channel $\Delta_{\lambda}$, defined as follows:
\be
\Delta_{\lambda}(M) = \lambda M + \frac{1 - \lambda}{2} \, \tr (M) \, I_2 \qquad M \in {\cal M}_2
\ee
Complete positivity requires that $-1/3 \le \lambda \le 1$ \cite{KR1}.
For $\lambda > 0$ we can write $\lambda = e^{-t}$, and then
\be
\Delta_{\lambda} = e^{- t H_u}
\ee
where $H_u$ is the `uniform' generator given by restricting the identity map ${\cal I}_2$ to the traceless matrices:
\be\label{def:Hu}
H_u(M) = M - \frac{1}{2} \, \tr (M) \, I_2
\ee
Clearly $\| H_u \|_{\min} = 1$, so $H_u \in G^{CP}_1$.
The hypercontractivity bound for $\Delta_{\lambda}$ is an immediate corollary of Theorem \ref{thm:main}.

\begin{cor}\label{cor:depol}
For all $1 < p \le q$ and all $n \ge 1$,
\be\label{depol1}
||| \Delta_{e^{-t}}^{\ot n} |||_{p \rightarrow q} =1 \,\, \mbox{if and only if} \,\, e^{-t} \le \sqrt{\frac{p-1}{q-1}}
\ee
\end{cor}

\medskip
\noindent \underline{\em Remarks.}
\medskip
\par\noindent{\em 1)}
Montanaro and Osborne \cite{MO1} proved the hypercontractive bound (\ref{depol1}) for the
qubit depolarizing channel 
for values of $p$ and $q$ restricted to the  range $1 < p \le 2 \le q$. Some applications of Corollary \ref{cor:depol} 
in quantum information theory are described in \cite{KT}, \cite{MO1} and \cite{M1}.

\medskip
\par\noindent{\em 2)}
The norm $||| e^{ - t_1 H_1} \ot \cdots \ot e^{- t_n H_n} |||_{p \rightarrow q}$
is always bounded below by $1$. Thus the non-trivial content of Theorem \ref{thm:main} is the upper bound.
Furthermore since $e^{ - t_1 H_1} \ot \cdots \ot e^{- t_n H_n}$ is completely positive it follows \cite{Wat}, \cite{Au1} that the supremum in
(\ref{def:map-norm}) is always achieved on positive semidefinite matrices. Thus the `if' part of the Theorem is a consequence of the following
bound (also making use of (\ref{norms-relate})): for all $A \in {\cal M}_{2^n}$ with $A \ge 0$, for all $1 < p \le q$ and all $n \ge 1$
\be\label{main2}
2^{-n/q} \, || e^{ - t_1 H_1} \ot \cdots \ot e^{- t_n H_n}(A) ||_q \le 2^{-n/p} \, || A ||_p \,\,\,\, \mbox{if} \,\,\,\, \max_i e^{- t_i} \le \sqrt{\frac{p-1}{q-1}}
\ee

\medskip
\par\noindent{\em 3)}
For matrices which are diagonal in the computational basis, the norms in (\ref{main1}) reduce
to classical $l^p$ norms of functions. The computational basis consists of products of diagonal matrices
$\{E_{s_1} \ot \cdots \ot E_{s_n}\}$ where $s=(s_1,\dots,s_n) \in \{0,1\}^n$, and 
\be
E_0 = \bmx {1 & 0 \cr 0 & 0 } \emx, \, E_1 = \bmx {0 & 0 \cr 0 & 1 } \emx
\ee
Thus if we define ${\cal D} \subset {\cal M}_{2^n}$ to be the subalgebra of diagonal matrices of the form
\be\label{def:D}
D = \sum_{s_1,\dots,s_n \in \{0,1\}} f(s_1,\dots,s_n) \, E_{s_1} \ot \cdots \ot E_{s_n}
\ee
then the  norm of $D \in {\cal D}$ is
\be
|| D ||_p = \left( \sum_{s_1,\dots,s_n = 0,1} \, |f(s_1,\dots,s_n)|^p \right)^{1/p} = || f ||_p
\ee
There is a $1-1$ correspondence between matrices in ${\cal D}$ and complex-valued functions 
on the `boolean cube' $f : \{ 0,1\}^n \rightarrow \mathbb{C}$. Furthermore
the action of the depolarizing channel $\Delta_{\lambda}^{\ot n}$ on ${\cal D}$ corresponds to the 
action of the noise operator $T_{\lambda}$ on the coefficients. This noise operator is defined by
\be\label{def:T}
(T_{\lambda} f)(s) = \mathbb{E}_{s' \sim s} [f(s')]
\ee
where the expectation is taken over binary strings $s'$ obtained from $s$ by flipping each bit independently with
probability $(1 - \lambda)/2$ and leaving the bit unchanged with probability $(1+\lambda)/2$.
It follows that for $D$ of the form (\ref{def:D}),
\be
\Delta_{\lambda}^{\ot n}(D) = \sum_{s_1,\dots,s_n = 0,1} (T_{\lambda} f)(s_1,\dots,s_n) \, E_{s_1} \ot \cdots \ot E_{s_n}
\ee
and thus
\be
||| \Delta_{\lambda}^{\ot n} |||_{p \rightarrow q} = ||| T_{\lambda} |||_{p \rightarrow q}
\ee
The hypercontractivity bound for $T_{\lambda}$ was derived in
\cite{Bon} and \cite{Gr1}: writing $\lambda = e^{-t}$, the result is
that for all $1 < p \le q$ and all $n \ge 1$,
\be\label{BG1}
||| T_{e^{-t}} |||_{p \rightarrow q} =1 \,\, \mbox{if and only if} \,\, e^{-t} \le \sqrt{\frac{p-1}{q-1}}
\ee
In the case $n=1$ this result is known as $2$-point hypercontractivity, so we will refer to the
result (\ref{BG1}) as $2^n$-point hypercontractivity.
In some sense Theorem \ref{thm:main} can be seen as a kind of `non-commutative extension'
of this classical hypercontractivity bound.

\medskip
\par\noindent{\em 4)}
Theorem \ref{thm:main} is closely related to hypercontractivity for fermions. 
Recall the definitions of the Pauli matrices:
\be\label{def:Pauli}
\sigma_0 = \bmx {1 & 0 \cr 0 & 1} \emx, \,
\sigma_1 = \bmx {0 & 1 \cr 1 & 0} \emx, \,
\sigma_2 = \bmx {0 & -i \cr i & 0} \emx, \,
\sigma_3 = \bmx {1 & 0 \cr 0 & -1} \emx
\ee
(notice that we define $\sigma_0 = I_2$ to be the identity matrix.) Let
\be
y_k = \sigma_3 \ot \cdots \ot \sigma_3 \ot \sigma_1 \ot \sigma_0 \ot \cdots \ot \sigma_0, \quad k=1,\dots,n
\ee
where in $y_k$ the factor $\sigma_1$ appears in the $k^{th}$ position. The matrices $\{y_1,\dots,y_n\}$ generate 
a representation $\cal C$ of the $n$-dimensional Clifford algebra in ${\cal M}_{2^n}$. This algebra $\cal C$ can be identified with the algebra of 
configuration observables for a system of $n$ fermions, and the fermion oscillator semigroup acts as a contraction $P_t$
on $\cal C$. Following earlier ideas and results of Gross \cite{Gr2},
Carlen and Lieb \cite{CL1} proved the optimal hypercontractivity bound for 
the operator $P_t$ applied to matrices in $\cal C$.
The Carlen-Lieb result can be formulated in the setting of Theorem \ref{thm:main}.
Recall that the phase-damping channel $\gamma_{\lambda}$ for qubits is defined by
\be
\gamma_{\lambda}(M) = \frac{1+\lambda}{2} \, M + \frac{1 - \lambda}{2} \, \sigma_3 M \sigma_3, \quad
-1 \le \lambda \le 1
\ee
For $\lambda > 0$ the phase-damping channel belongs to a semigroup: 
$\gamma_{e^{-t}} = e^{ - t \Gamma_3}$ where the generator is defined by 
\be
\Gamma_3(M) = \frac{1}{2} \, M - \frac{1}{2} \, \sigma_3 M \sigma_3
\ee
When restricted to $\cal C$, $\gamma_{\lambda}^{\ot n}$ coincides with the oscillator semigroup
$P_t$ from  \cite{CL1}, with the identification $\lambda = e^{-t}$.
Thus the `if' part of the Carlen-Lieb bound can be re-stated as follows: for all $C \in {\cal C}$,
for all $1 < p \le q$ and all $n \ge 1$
\be\label{main4}
2^{-n/q} \, || \gamma_{e^{-t}}^{\ot n}(C) ||_q \le 2^{-n/p} \, || C ||_p \,\,\,\, \mbox{if} \,\,\,\, e^{-t} \le \sqrt{\frac{p-1}{q-1}}
\ee
Note however that (\ref{main4}) does {\em not} follow from Theorem \ref{thm:main},
as $\| \Gamma_3 \|_{\min} = 0$.

\medskip
\par\noindent{\em 5)}
Biane \cite{Bi} extended the Carlen-Lieb result to other subalgebras of ${\cal M}_{2^n}$.
Namely, consider any function $\epsilon : \{1,\dots,n\} \times \{1,\dots,n\} \rightarrow \{0,3\}$, and define the 
matrices
\be
z_k =  \sigma_{\epsilon(1,k)} \ot \cdots \ot \sigma_{\epsilon(k-1,k)} \ot \sigma_1 \ot \sigma_0 \ot \cdots \ot \sigma_0
\ee
where again the factor $\sigma_1$ appears in the $k^{th}$ position. When restricted to the algebra
generated by $\{z_1,\dots,z_n\}$, $\gamma_{e^{-t}}^{\ot n}$ again coincides with the number operator
$P_t^{\epsilon}$ defined by Biane \cite{Bi}.
Biane proved the hypercontractivity bound for $P_t^{\epsilon}$ restricted to the algebra generated by $(z_1,\dots,z_n)$. 
Note again that this result does not follow from Theorem \ref{thm:main} (except in the case $\epsilon(i,j) \equiv 0$,
in which case it reduces to the classical $2^n$-point hypercontractivity result).

 \medskip
\par\noindent{\em 6)}
The result of Theorem \ref{thm:main} is equivalent to a certain form of multiplicativity for products
of qubit  channel semigroups.
If we write $\Phi_j = e^{ - t_j H_j}$ then
\bee
||| \Phi_1 \ot \cdots \ot \Phi_n |||_{p \rightarrow q} 
&=& 2^{-n/q + n/p} \,|| \Phi_1 \ot \cdots \ot \Phi_n ||_{p \rightarrow q} \\
& \ge & 2^{-n/q + n/p} \,\left(|| \Phi_1 ||_{p \rightarrow q}\right) \dots \left(|| \Phi_n ||_{p \rightarrow q}\right) \\
& \ge & 2^{-n/q + n/p} \, \left( \frac{|| \Phi(I_2) ||_q}{|| I_2 ||_p} \right)^n \\
& = & 1
\eee
where the first inequality follows by restricting to product states, the second by evaluating on the identity
matrix, and the last equality holds because the channels are unital.
Thus  the fact that the hypercontractivity bound (\ref{main1}) holds at some values $p,q$ is equivalent to two conditions: 
first, the $p \rightarrow q$ norm for $\Phi_1 \ot \cdots \ot \Phi_n$ is multiplicative, and second the $p \rightarrow q$ norm 
for each individual channel $\Phi_j$ is achieved at the identity matrix. Other results are known for these channels;
for example the $1 \rightarrow q$ norm 
is multiplicative for any product of unital qubit channels \cite{Ki3}, and this norm is achieved on a pure state. In this paper we will 
prove a  result which settles the multiplicativity question for a few more cases (see Theorem \ref{thm:mult} below),
namely that the $p \rightarrow q$ norm is multiplicative for any product of unital qubit channels under the condition $1 \le p \le 2 \le q$.

 \medskip
\par\noindent{\em 7)}
Not all unital qubit channels belong to self-adjoint semigroups. 
One such example is the two-Pauli channel of Bennett, Fuchs and Smolin
\cite{BFS} which has the following form:
\be
\Theta_{\lambda}(M) = \lambda M + \frac{1 - \lambda}{2} \sigma_1 M \sigma_1
+\frac{1-\lambda}{2} \sigma_2 M \sigma_2, \qquad
0 \le \lambda \le 1
\ee
It is an interesting problem to determine if there are bounds analagous to Theorem \ref{thm:main}
for such channels.

\medskip
The paper is organized as follows. In Section \ref{sect:mult} we review some results for qubit channels,
and prove a representation theorem for self-adjoint generators of unital qubit semigroups.
We also derive several lemmas needed for the proof of Theorem \ref{thm:main}.
One result (Theorem \ref{thm:mult} ) proves multiplicativity  for the $p \rightarrow q$ norm of a product $\Omega \ot \Phi$,
where $\Omega$ is any completely positive map and $\Phi$ is a unital qubit channel. We also derive
a version of Gross' Lemma \cite{Gr2} which is adapted to matrix algebras. In Section
\ref{sect:proof} we prove Theorem \ref{thm:main}, and along the way we derive  a logarithmic
Sobolev inequality for qubit semigroups.

\section{Unital qubit channels, a multiplicativity result, and Gross' Lemma}\label{sect:mult}
A qubit channel $\Phi$ can be conveniently represented using the Pauli matrix basis defined in (\ref{def:Pauli}). Every matrix in ${\cal M}_2$ can be written in the form
$\sum_{i=0}^3 a_i \, \sigma_i$
for some complex coefficients $\{a_i\}$. With respect to this basis $\Phi$ is then
represented by a $4 \times 4$ matrix, with entries
\be
\Phi_{i j} = \frac{1}{2} \, \tr \sigma_i \, \Phi(\sigma_j), \quad i,j = 0,1,2,3
\ee
Since $\Phi$ preserves the subspace of self-adjoint matrices in ${\cal M}_2$, this matrix is real \cite{KR1}.
If in addition $\Phi$ is unital, meaning that $\Phi(I_2) = I_2$ then
by making suitable unitary transformations in the domain and range if necessary, $\Phi$ can be put into
diagonal form with respect to this basis \cite{KR1}, \cite{RSW}:
\be\label{diag-form}
\Phi(\sigma_i) = \lambda_i \, \sigma_i, \quad i=0,1,2,3
\ee
where $\lambda_0 =1$ and the other three parameters satisfy the conditions required for complete positivity:
\be\label{conds-CP}
\lambda_1 + \lambda_2 - \lambda_3 &\le & 1 \nonumber \\
\lambda_1 - \lambda_2 + \lambda_3 &\le & 1 \nonumber \\
- \lambda_1 + \lambda_2 + \lambda_3 &\le & 1 \nonumber \\
- \lambda_1 - \lambda_2 - \lambda_3 &\le & 1
\ee
As a shorthand we will write $\Phi = (\lambda_1,\lambda_2,\lambda_3)$ to indicate that $\Phi$ is diagonal and
satisfies (\ref{diag-form}).

\medskip
The proof of Theorem  \ref{thm:main} will use the following multiplicativity result for
unital qubit channels.

\begin{thm}\label{thm:mult}
Let $\Phi$ be a unital qubit channel and let $\Omega$ be a completely positive map. Then for
all $1 \le p \le 2 \le q$,
\be\label{mult1}
|| \Omega \ot \Phi ||_{p \rightarrow q} = || \Omega ||_{p \rightarrow q} \, || \Phi ||_{p \rightarrow q}
\ee
\end{thm}

\medskip
\par\noindent{\em Proof:}
by restricting to product states, it follows immediately that the left side of (\ref{mult1}) is bounded below by the
right side. So we must show that the left side is bounded above by the right side. First, by unitary invariance
of the norm we may assume without loss of generality that $\Phi$ is a diagonal qubit channel. Secondly,
since $\Omega \ot \Phi$
is completely positive, its ${p \rightarrow q}$ norm is achieved on a positive semidefinite matrix
\cite{Wat}, \cite{Au1}, and thus it is sufficient to prove that for all $\Phi = (\lambda_1,\lambda_2,\lambda_3)$ and all $A \ge 0$
\be\label{mult2}
|| (\Omega \ot \Phi)(A) ||_q \le || \Omega ||_{p \rightarrow q} \, || \Phi ||_{p \rightarrow q} \, || A ||_p
\ee

\medskip
Now suppose that $\Omega$ is a completely positive map in ${\cal L}({\cal M}_k)$. It follows that 
$A \in {\cal M}_{k} \ot {\cal M}_2 = {\cal M}_{2 k}$, and so $A$ can be written as a
$2 \times 2$ block matrix:
\be
A = \bmx {A_{11} & A_{12} \cr A_{12}^* & A_{22}} \emx
\ee
where $A_{11},A_{12},A_{22} \in {\cal M}_k$.
Positivity requires that $A_{12} = A_{11}^{1/2} R A_{22}^{1/2}$ for some contraction $R$. Define
\be
B_{11} = \Omega(A_{11}), \quad B_{12} = \Omega(A_{12}), \quad B_{22} = \Omega(A_{22})
\ee
and
\be\label{mult-ineq2}
C_{11} &=& \frac{1 + \lambda_3}{2} B_{11} + \frac{1 - \lambda_3}{2} B_{22} \nonumber \\
C_{12} &=& \frac{\lambda_1 + \lambda_2}{2} B_{12} + \frac{\lambda_1 - \lambda_2}{2} B_{12}^* \nonumber \\
C_{22} &=& \frac{1 - \lambda_3}{2} B_{11} + \frac{1 + \lambda_3}{2} B_{22}
\ee
Then it follows that
\bee
(\Omega \ot \Phi)(A) &=& ({\cal I}_k \ot \Phi) \bmx {B_{11} & B_{12} \cr B_{12}^* & B_{22}} \emx \\
&=& \bmx {C_{11} & C_{12} \cr C_{12}^* & C_{22}} \emx \\
\eee
We now follow the strategy used in \cite{MO1} by applying Theorem 1 in \cite{Ki2}: for $q \ge 2$ this gives
\be\label{mult-ineq3}
|| (\Omega \ot \Phi)(A) ||_q &=& \left\| \bmx {C_{11} & C_{12} \cr C_{12}^* & C_{22}} \emx \right\|_q  \nonumber \\
& \le & \left\| \bmx {|| C_{11} ||_q & || C_{12} ||_q \cr || C_{12} ||_q & || C_{22} ||_q} \emx \right\|_q
\ee
From (\ref{mult-ineq2}) we deduce
\be\label{mult-ineq3a}
|| C_{11} ||_q & \le & \frac{1 + \lambda_3}{2} || B_{11} ||_q + \frac{1 - \lambda_3}{2} || B_{22} ||_q \nonumber \\
|| C_{12} ||_q & \le & \max \{|\lambda_1|, |\lambda_2| \} \, || B_{12} ||_q \nonumber \\
|| C_{22} ||_q & \le & \frac{1 - \lambda_3}{2} || B_{11} ||_q + \frac{1 + \lambda_3}{2} || B_{22} ||_q
\ee
The $2 \times 2$ matrix in the last line of (\ref{mult-ineq3}) is positive semidefinite, and thus its norm
is an increasing function of its diagonal entries. Therefore
\be\label{mult-ineq4}
& \hskip-3.5in \left\| \bmx {|| C_{11} ||_q & || C_{12} ||_q \cr || C_{12} ||_q & || C_{22} ||_q} \emx \right\|_q & \nonumber \\
\le
& \left\| \bmx {\frac{1 + \lambda_3}{2} || B_{11} ||_q + \frac{1 - \lambda_3}{2} || B_{22} ||_q & || C_{12} ||_q \cr || C_{12} ||_q & \frac{1 - \lambda_3}{2} || B_{11} ||_q + \frac{1 + \lambda_3}{2} || B_{22} ||_q} \emx \right\|_q &
\ee
The matrix on the right side of (\ref{mult-ineq4}) is positive semidefinite. Letting $t$ denote its off-diagonal entry, it follows that the norm of the matrix is an increasing function of $t$ in the interval $0 \le t \le t_m$, where $t_m$ is the value at which the matrix becomes singular. In our case we wish to replace the entry $|| C_{12} ||_q$ by the second bound in (\ref{mult-ineq3a}) and deduce that the norm increases, so we must check that the resulting matrix after the replacement is positive semidefinite. Indeed $ \left( \max \{|\lambda_1|, |\lambda_2| \} \, || B_{12} ||_q \right)^2 \le \left( || B_{12} ||_q \right)^2 \le || B_{11} ||_q \, || B_{22} ||_q$, where the second inequality is a consequence of the positivity of the matrix $(\Omega \ot {\cal I}_2)(A)$. Furthermore
\bee
&  & \hskip-0.4in || B_{11} ||_q \, || B_{22} ||_q  \\
& = & \frac{1 - \lambda_3^2}{2} || B_{11} ||_q \, || B_{22} ||_q + \frac{1 + \lambda_3^2}{2} || B_{11} ||_q \, || B_{22} ||_q  \\
& \le & \frac{1 - \lambda_3^2}{4} \left(|| B_{11} ||_q^2 + || B_{22} ||_q^2 \right) 
+ \frac{1 + \lambda_3^2}{2} || B_{11} ||_q \, || B_{22} ||_q  \\
& = & \left( \frac{1 + \lambda_3}{2} || B_{11} ||_q + \frac{1 - \lambda_3}{2} || B_{22} ||_q \right) \,
\left(\frac{1 - \lambda_3}{2} || B_{11} ||_q + \frac{1 + \lambda_3}{2} || B_{22} ||_q \right) 
\eee
Thus the norm does increase when the off-diagonal entry $|| C_{12} ||_q$ is replaced by the right side of the bound in (\ref{mult-ineq3a}), and hence
\be\label{mult-ineq5}
& \hskip-3.5in  || (\Omega \ot \Phi)(A) ||_q & \nonumber \\
\le 
& \left\| \bmx {\frac{1 + \lambda_3}{2} || B_{11} ||_q + \frac{1 - \lambda_3}{2} || B_{22} ||_q 
& \max \{|\lambda_1|, |\lambda_2| \} \, || B_{12} ||_q \cr 
\max \{|\lambda_1|, |\lambda_2| \} \, || B_{12} ||_q & \frac{1 - \lambda_3}{2} || B_{11} ||_q + \frac{1 + \lambda_3}{2} || B_{22} ||_q} \emx \right\|_q &
\ee
Define
\be
\mu = \begin{cases} {\lambda_1 & if $|\lambda_1| \ge |\lambda_2|$ \cr
\lambda_2 & if $|\lambda_1| < |\lambda_2|$ } \end{cases}, \qquad
\epsilon = \begin{cases} {1 & if $|\lambda_1| \ge |\lambda_2|$ \cr
i & if $|\lambda_1| < |\lambda_2|$} \end{cases}
\ee
Since $| \epsilon | = 1$, we can use unitary invariance of the norm to re-write the right side of (\ref{mult-ineq5}) as
\be\label{mult-ineq7}
&  
 \left\| \bmx {\frac{1 + \lambda_3}{2} || B_{11} ||_q + \frac{1 - \lambda_3}{2} || B_{22} ||_q 
& \mu \,\epsilon \, || B_{12} ||_q \cr 
\mu \,\overline{\epsilon} \, \, || B_{12} ||_q & \frac{1 - \lambda_3}{2} || B_{11} ||_q + \frac{1 + \lambda_3}{2} || B_{22} ||_q} \emx \right\|_q & \nonumber \\
& =
\left\| \Phi  \bmx {|| B_{11} ||_q & \epsilon || B_{12} ||_q \cr \overline{\epsilon} || B_{12} ||_q & || B_{22} ||_q} \emx \right\|_q & 
\ee
Using (\ref{mult-ineq7}) and the definition of $|| \Phi ||_{p \rightarrow q}$ on the right side of 
the inequality (\ref{mult-ineq5}) we obtain
\be\label{mult-ineq6}
|| (\Omega \ot \Phi)(A) ||_q 
& \le & || \Phi ||_{p \rightarrow q} \,  \left\| \bmx {|| B_{11} ||_q & \epsilon || B_{12} ||_q \cr \overline{\epsilon} || B_{12} ||_q & || B_{22} ||_q} \emx \right\|_p \nonumber \\
&=& || \Phi ||_{p \rightarrow q} \,  \left\| \bmx {|| B_{11} ||_q &  || B_{12} ||_q \cr  || B_{12} ||_q & || B_{22} ||_q} \emx \right\|_p
\ee
where we again used unitary invariance of the norm in the last equality.
We also have
\be
|| B_{ij} ||_q \le || \Omega ||_{p \rightarrow q} \, || A_{ij} ||_p, \quad i,j = 1,2
\ee
and by the same argument as before we can deduce that the norm on the last line of (\ref{mult-ineq6})
increases when these bounds are applied to the entries of the matrix. Thus
\be\label{mult-ineq7}
|| (\Omega \ot \Phi)(A) ||_q \le
 || \Phi ||_{p \rightarrow q} \,  || \Omega ||_{p \rightarrow q} \,
 \left\| \bmx {|| A_{11} ||_p &  || A_{12} ||_p \cr  || A_{12} ||_p & || A_{22} ||_p} \emx \right\|_p
\ee
Finally we again use the inequality from Theorem 1 in \cite{Ki2}, this time for $p \le 2$, and deduce that
\be\label{mult-ineq8}
|| (\Omega \ot \Phi)(A) ||_q & \le&
 || \Phi ||_{p \rightarrow q} \,  || \Omega ||_{p \rightarrow q} \,
 \left\| \bmx {A_{11}  &   A_{12} \cr   A_{12}  &  A_{22}} \emx \right\|_p \nonumber \\
 &=& || \Phi ||_{p \rightarrow q} \,  || \Omega ||_{p \rightarrow q} \, || A ||_p
\ee

\noindent \qed

\medskip
Now consider a qubit channel generator $H \in G^{CP}$. Since by assumption the channel
$e^{ - tH}$ is completely positive for all $t \ge 0$, it follows that $e^{ - tH}$ is
represented by a real matrix with respect to the Pauli basis, and hence 
$H$ is also represented by a real matrix. Furthermore since $H = \widehat{H}$ this matrix is symmetric,
and thus there is an orthogonal matrix which diagonalizes $H$.
This orthogonal matrix is implemented by a unitary transformation on ${\cal M}_2$.
Since all matrix norms are unitarily invariant, we can without loss of generality assume 
 that $H$ is a diagonal matrix in the Pauli basis.
That is,
\be\label{diag-form2}
H(\sigma_0) = 0, \quad  H(\sigma_i) = h_i \, \sigma_i, \quad i=1,2,3
\ee
We will write $H=(h_1,h_2,h_3)$ to indicate that $H$ is diagonal and satisfies (\ref{diag-form2}).
It follows that $e^{- t H} = (e^{-t h_1}, e^{ - t h_2}, e^{ - t h_3})$ is also diagonal for all $t \ge 0$.
Also it is easy to see that in this case
\be
\| H \|_{\min} = \min \{h_1,h_2,h_3 \}
\ee

\medskip
The coefficients $h_i$ satisfy additional constraints, imposed by the conditions (\ref{conds-CP}).
Define three special generators:
\be\label{def:Gamma}
\Gamma_1 = (0,1,1),\quad
\Gamma_2 = (1,0,1),\quad
\Gamma_3 = (1,1,0)
\ee

\begin{lemma}\label{lem:span}
A diagonal generator $H$ is in $G^{CP}$ if and  only if there are constants  $a_1,a_2,a_3 \ge 0$ such that
\be\label{span-1}
H = a_1 \Gamma_1 + a_2 \Gamma_2 + a_3 \Gamma_3
\ee
\end{lemma}

\medskip
\par\noindent {\em Proof:}
suppose that $H  = (h_1,h_2,h_3) \in G^{CP}$ and
let
\bee
a_1 = \frac{1}{2} \, (- h_1 + h_2 + h_3), \quad
a_2 = \frac{1}{2} \, (h_1 - h_2 + h_3), \quad
a_3 = \frac{1}{2} \, (h_1 + h_2 - h_3)
\eee
then it is easily checked that $H = \sum_{i=1}^3 a_i \Gamma_i$.
Also let
\be\label{span-2}
\lambda_i = e^{ - t h_i}, \quad i=1,2,3
\ee
Then the condition that $e^{- t H}$ is CP for all $t \ge 0$ implies (by taking derivatives
of the conditions (\ref{conds-CP}) at $t=0$) that $a_i \ge 0$ for $i=1,2,3$. This proves the
`only if' part of the Lemma. For the `if' part, assume that (\ref{span-1}) holds with $a_i \ge 0$ and define
\bee
h_1 = a_2 + a_3, \quad
h_2 = a_1 + a_3, \quad
h_3 = a_1 + a_1
\eee
Then using the definitions in (\ref{span-2}) again, we find that
\bee
-\lambda_1 + \lambda_2 + \lambda_3 &\le&
- e^{-a_2 t} + e^{ - a_1 t} + e^{ - a_1 t} e^{ - a_2 t} \\
& \le & \begin{cases} {e^{ - a_1 t} e^{ - a_2 t} & if $a_2 \le a_1$ \cr
e^{ - a_1 t} - e^{-a_2 t} (1 - e^{ - a_1 t}) & if $a_2 > a_1$ } \end{cases} \\
& \le & 1
\eee
Similarly the other conditions in (\ref{conds-CP}) hold for all $t \ge 0$.
Thus $e^{- t H}$ is CP for all $t \ge 0$, and hence $H \in G^{CP}$.

\medskip
\par\noindent \qed

\medskip
The following inequality is a matrix algebra version of Gross' Lemma \cite{Gr2}. 
We include some of the details of the proof, and then refer to \cite{Gr2} for the rest.
Recall the notation $\langle A, B \rangle = \tr A^* B$ introduced in Section \ref{intro}.

\begin{lemma}[Gross 1975]\label{Gross1}
Let $H \in G^{CP}$, and for $n \ge 1$ let $H^{(n)} = {\cal I}_{2^{n-1}} \ot H$. Then for any $A \in {\cal M}_{2^n}$, $A \ge 0$,
and $p \ge 1$,
\be
\langle A^{p/2}, H^{(n)}(A^{p/2}) \rangle \le \frac{(p/2)^2}{p-1} \, \langle A, H^{(n)}(A^{p-1}) \rangle
\ee
\end{lemma}

\medskip
\par\noindent{\em Proof:}
by Lemma \ref{lem:span} it is enough to prove the result for $H = \Gamma_i$, $i=1,2,3$.
Furthermore the generators $\Gamma_i$ are all unitarily equivalent, so it is sufficient
to use $\Gamma_3$.
We write $A$ in block matrix form
\bee
A = \bmx {X & Y \cr Y^* & Z } \emx
\eee
and define
\bee
f(s) = \bmx {X & s Y \cr s Y^* & Z } \emx, \quad -1 \le s \le 1
\eee
Positivity of $A$ implies that $f(s) \ge 0$ for all $-1 \le s \le 1$, and
\bee
({\cal I}_{2^{n-1}} \ot \Gamma_3)(A) &=& \frac{1}{2} \left(A - ({\cal I}_{2^{n-1}} \ot \sigma_3) \, A \, ({\cal I}_{2^{n-1}} \ot \sigma_3) \right) \\
&=& \frac{1}{2} \left( f(1) - f(-1) \right)
\eee
Similarly 
\bee
({\cal I}_{2^{n-1}} \ot \Gamma_3)(A^{p/2}) &=&  \frac{1}{2} \left(f(1)^{p/2} - f(-1)^{p/2} \right) \\
&=& \frac{1}{2} \int_{-1}^1 \frac{d}{ds} f(s)^{p/2} \,  d s
\eee
Using the relation $({\cal I}_{2^{n-1}} \ot \Gamma_3)^2 = ({\cal I}_{2^{n-1}} \ot \Gamma_3)$
and the Cauchy-Schwarz inequality
we deduce
\bee
\langle A^{p/2}, ({\cal I}_{2^{n-1}} \ot \Gamma_3)(A^{p/2}) \rangle \le
  \frac{1}{2} \int_{-1}^1 \langle \frac{d}{ds} f(s)^{p/2},  \frac{d}{ds} f(s)^{p/2} \rangle \,  d s
\eee
Similarly
\bee
\langle A, ({\cal I}_{2^{n-1}} \ot \Gamma_3)(A^{p-1}) \rangle =
  \frac{1}{2} \int_{-1}^1 \langle \frac{d}{ds} f(s),  \frac{d}{ds} f(s)^{p-1} \rangle \,  d s
\eee
Thus it is sufficient to prove that for $|s| \le 1$,
\be\label{Gr-pf4}
\langle \frac{d}{ds} f(s)^{p/2},  \frac{d}{ds} f(s)^{p/2} \rangle \le
\frac{(p/2)^2}{p-1} \, 
 \langle \frac{d}{ds} f(s),  \frac{d}{ds} f(s)^{p-1} \rangle
 \ee
As noted in Gross' proof, it is sufficient to prove the bound (\ref{Gr-pf4}) for
$p = 2m/n$ where $m,n$ are integers with $1 \le n < m$.
Letting $g(s) = f(s)^{1/n}$, the proof reduces to the inequality:
\be\label{Gr-pf5}
\langle \frac{d}{ds} g(s)^{m},  \frac{d}{ds} g(s)^{m} \rangle \le
\frac{(p/2)^2}{p-1} \, 
 \langle \frac{d}{ds} g(s)^n,  \frac{d}{ds} g(s)^{2m - n} \rangle
 \ee
Gross uses a combinatorial argument to prove (\ref{Gr-pf5}), in the case where
$g$ takes (positive) values in the Clifford algebra. The proof goes over word for word to the present
case where $g$ is a positive matrix-valued function.

\par\noindent \qed

\medskip
As an immediate corollary of Lemma \ref{Gross1} we deduce that certain norms are monotone.

\begin{cor}\label{cor:non-inc1}
Let $H \in G^{CP}$, and for $n \ge 1$ let $H^{(n)} = {\cal I}_{2^{n-1}} \ot H$. 
Then for any $A \ge 0$, and any $q \ge 1$ the function
\be\label{non-inc1}
t \, \mapsto \| ({\cal I}_{2^{n-1}} \ot e^{ - t H})(A) \|_{q}
\ee
is non-increasing.
\end{cor}

\medskip
\par\noindent{\em Proof:}
let 
\bee
B = ({\cal I}_{2^{n-1}} \ot e^{ - t H})(A)^q
\eee
and
\bee
f(t) = \ln \| B \|_{q}
\eee
Then 
\bee
f'(t) = - \frac{1}{\tr B^q} \, \tr \left( B^{q-1} H^{(n)}(B) \right)
\eee
By assumption $H$  is self-adjoint and $B \ge 0$, thus
\bee
\tr B^{q-1} H^{(n)}(B) = \tr B H^{(n)}(B^{q-1}) = \langle B, H^{(n)}(B^{q-1}) \rangle
\eee
Applying Lemma \ref{Gross1} we deduce that
\bee
f'(t) \le - \frac{1}{\tr B^q} \,\frac{q-1}{(q/2)^2} \, \langle B^{q/2}, H^{(n)}(B^{q/2}) \rangle) \le 0
\eee
since $H \ge 0$ implies that $H^{(n)} \ge 0$.
Hence $\| B \|_q$ is non-increasing as a function of $t$.
\par\noindent\qed

\section{Proof of Theorem \ref{thm:main}}\label{sect:proof}
As discussed at the start of Section \ref{sect:mult}, the generators $H_j$ may be diagonalized
using unitary transformations. Furthermore the channels $(h_1,h_2,h_3)$ and 
$(h_2,h_3,h_1)$ are unitarily equivalent for any $h_1,h_2,h_3$.
Since $\| H_j \|_{\min} = 1$ we can therefore assume
without loss of generality  that 
\be\label{assume-H}
H_j = (h_{j,1},h_{j,2},1) \quad \mbox{with} \quad h_{j,1},h_{j,2} \ge 1
\ee

\medskip
We first establish the `only if' part of the result. To this end, suppose that for some $j$ we have
\be\label{cond-onlyif}
e^{- t_j} > \sqrt{\frac{p-1}{q-1}}
\ee
As shown in Remark 6, we have the lower bound
\be\label{onlyif3}
||| \Phi_1 \ot \cdots \ot \Phi_n |||_{p \rightarrow q}
& \ge & \left(||| \Phi_1 |||_{p \rightarrow q}\right) \dots \left(||| \Phi_n |||_{p \rightarrow q}\right) \nonumber \\
& \ge & ||| \Phi_j |||_{p \rightarrow q}
\ee
where $\Phi_i = e^{- t_i H_i}$, $i=1,\dots,n$.
Furthermore
\bee
||| \Phi_j |||_{p \rightarrow q} \ge
\sup_{M \, {\rm diagonal}}  \frac{||| \Phi_j(M) |||_q}{||| M |||_p} 
\eee
where the supremum is restricted to diagonal matrices. For any diagonal matrix $M$, (\ref{assume-H}) implies that
\bee
\Phi_j (M) = \Delta_{e^{-t_j}}(M)
\eee
where $\Delta$ is the depolarizing qubit channel.
Thus
\bee
||| \Phi_j |||_{p \rightarrow q} \ge
\sup_{M \, {\rm diagonal}}  \frac{||| \Delta_{e^{-t_j}}(M) |||_q}{||| M |||_p} 
\eee
As discussed in Remark 3, the action of
$\Delta_{e^{-t_j}}$ on diagonal matrices is equivalent to the action of the noise operator $T_{e^{-t_j}}$ on functions
$f \,:\, \{0,1\} \rightarrow \mathbb{C}$. Thus the classical $2$-point
hypercontractivity result applies, and from the condition (\ref{cond-onlyif}) we conclude that
\bee
\sup_{M \, {\rm diagonal}}  \frac{||| \Delta_{e^{-t_j}}(M) |||_q}{||| M |||_p} > 1
\eee
Therefore $||| \Phi_j |||_{p \rightarrow q} > 1$, and from (\ref{onlyif3}) this implies 
\bee
||| \Phi_1 \ot \cdots \ot \Phi_n |||_{p \rightarrow q} > 1
\eee

\medskip
Now we turn to the proof of the `if' part of Theorem \ref{thm:main}.
As explained in Remark 2, this is equivalent to the following bound:
for all $A \ge 0$, for all $1 < p \le q$ and all $n \ge 1$
\be\label{main-pf1}
2^{-n/q} \, || (e^{ - t_1 H_1} \ot \cdots \ot e^{- t_n H_n})(A) ||_q \le 2^{-n/p} \, || A ||_p \,\,\,\, \mbox{if} \,\,\,\, \max_j e^{- t_j} \le \sqrt{\frac{p-1}{q-1}}
\ee
First we apply Corollary \ref{cor:non-inc1} to conclude that the left side of
(\ref{main-pf1}) is non-increasing as a function of $t_j$, for all $j=1,\dots,n$.
Thus it is sufficient to assume that
\bee
e^{- t_j} = \sqrt{\frac{p-1}{q-1}}, \quad j=1,\dots,n
\eee

\medskip
We will follow Gross' strategy of proof \cite{Gr2}, by first deriving a logarithmic Sobolev inequality, and then use this to prove
monotonicity of the left side of (\ref{main-pf1}) along a suitable curve $q(t)$. 
In order to derive the log-Sobolev inequality, we apply Lemma \ref{thm:mult} with $p=2$ and $q = 1 + e^{2 t}$.
Repeated application of Lemma \ref{thm:mult} shows that for all $A \ge 0$, and all $t \ge 0$,
\be\label{main-pf3}
|| (e^{ - t H_1} \ot \cdots \ot e^{- t H_n})(A) ||_{1 + e^{2 t}} \le 
\left( \prod_{j=1}^n \| e^{- t H_j} \|_{2 \rightarrow 1 + e^{2 t}} \right) \, 
 || A ||_2
\ee
Let $H_j = (h_1,h_2,h_3)$ with $h_1,h_2 \ge1$ and $h_3=1$.
Any positive matrix $C \in {\cal M}_2$ can be written as $C = c_0 \sigma_0 + \sum_{i=1}^3 c_i \sigma_i$
with $c_0 \ge 0$ and $ \sum_{i=1}^3 c_i^2 \le c_0^2$.
The eigenvalues of $C$ are
\bee
c_0 + \left( \sum_{i=1}^3 c_i^2 \right)^{1/2} \quad \mbox{and} \quad
c_0 - \left( \sum_{i=1}^3 c_i^2 \right)^{1/2}
\eee
Therefore
\bee
\| e^{- t H_j}(C) \|_q^q &=& \left(c_0 + \left( \sum_{i=1}^3 c_i^2 \, e^{- 2 t h_i}\right)^{1/2} \right)^q +
\left(c_0 - \left( \sum_{i=1}^3 c_i^2 \, e^{- 2 t h_i}\right)^{1/2} \right)^q \nonumber \\
& \le & \left(c_0 + e^{-t} \left( \sum_{i=1}^3 c_i^2 \right)^{1/2} \right)^q +
\left(c_0 - e^{-t} \left( \sum_{i=1}^3 c_i^2 \right)^{1/2} \right)^q \nonumber \\
&=& \| \Delta_{e^{-t}}(\tilde{C}) \|_q^q
\eee
where $\tilde{C}$ is the diagonal matrix
\bee
\tilde{C} = c_0 \sigma_0 + \left( \sum_{i=1}^3 c_i^2 \right)^{1/2} \sigma_3
\eee
Again we note that when restricted to diagonal matrices the operator $\Delta_{e^{-t}}$ coincides with
the classical noise operator.
Thus $2$-point hypercontractivity implies
\bee
\| e^{- t H_j}(C) \|_{1 + e^{2 t}} \le  \| \Delta_{e^{-t}}(\tilde{C}) \|_{1 + e^{2 t}} \le 
2^{1/(1 + e^{2t}) - 1/2} \, \| \tilde{C} \|_2 =
2^{1/(1 + e^{2t}) - 1/2} \, \| C \|_2 \nonumber
\eee
and hence
\bee
\| e^{- t H_j} \|_{2 \rightarrow 1 + e^{2 t}} \le 2^{1/(1 + e^{2t}) - 1/2}
\eee
Applying this bound to each term in the product on the right side of (\ref{main-pf3}) we deduce that
\be\label{main-pf4}
|| (e^{ - t H_1} \ot \cdots \ot e^{- t H_n})(A) ||_{1 + e^{2 t}} \le \left( 2^{1/(1 + e^{2t}) - 1/2} \right)^n \, 
 || A ||_2
\ee
and hence
\be\label{main-pf5}
2^{-n/(1 + e^{2t})} || (e^{ - t H_1} \ot \cdots \ot e^{- t H_n})(A) ||_{1 + e^{2 t}} \le 2^{-n/2} \, \| A \|_2
\ee
The inequality (\ref{main-pf5}) holds for all $t \ge 0$, with equality at $t=0$, hence the derivative of the
left side at $t=0$ must be non-positive. Computing this derivative produces the inequality
\be\label{main-pf6}
n \ln 2 - \ln \tr A^2 + \frac{1}{\tr A^2} \,
\tr \left[ A^2 \ln A^2 + 2 A \left( - \sum_{k=1}^n H^{(k)} \right)(A) \right] \le 0
\ee
where
\be\label{main-pf7}
H^{(k)} = {\cal I}_{2^{k-1}} \ot H_k \ot {\cal I}_{2^{n-k}}, \quad k=1,\dots,n
\ee
Using the normalized trace $\tau$ this can be written in the more standard form for a logarithmic Sobolev
inequality: for all $A \ge 0$
\be\label{main-pf8}
- \tau(A^2) \ln \tau (A^2) + \tau (A^2 \ln A^2) \le 2 \, \sum_{k=1}^n \tau\left( A H^{(k)}(A) \right)
\ee
The inequality (\ref{main-pf8}) was derived by Kastoryano and Temme \cite{KT} 
for the $n$-fold product of the qubit depolarizing channel,
which is obtained by setting $H_k = H_u$ (defined in (\ref{def:Hu})) for all $k$. In fact (\ref{main-pf8}) then follows from the 
bound for the qubit depolarizing channel, because the condition $\| H_k \|_{\min}=1$ for a generator $H_k$ implies that
\bee
\tau\left( A \,\, H^{(k)}(A) \right) \ge \tau\left( A \,\, ({\cal I}_{2^{k-1}} \ot H_u \ot {\cal I}_{2^{n-k}})(A) \right)
\eee
for any $A \ge 0$.

\medskip
Continuing with the proof of Theorem \ref{thm:main}, we will use (\ref{main-pf6}) to derive a monotonicity
result for the left side of (\ref{main-pf1}).
For $p \ge 1$ let $q(t) = 1 + e^{2 t} (p-1)$. We wish to prove that for all $A \ge 0$
\be\label{main-pf10}
2^{-n/q(t)} \, || (e^{ - t H_1} \ot \cdots \ot e^{- t H_n})(A) ||_{q(t)} \le 2^{-n/p} \, || A ||_p
\ee
Since equality holds at $t=0$ it is sufficient to prove that the left side of (\ref{main-pf10}) is a non-increasing
function of $t$, for all $t \ge 0$. Let
\be\label{main-pf10a}
B = (e^{ - t H_1} \ot \cdots \ot e^{- t H_n})(A), \quad
g(t) = \ln \left( 2^{-n/q} \, || B ||_q \right)
\ee
then we find
\be\label{main-pf11}
g'(t) = \frac{2}{\tr B^q} \, \frac{q-1}{q^2} \, \bigg[ && \hskip-0.35in
n \ln2 \, \tr B^q - \tr B^q \, \ln \tr B^q + \tr \left( B^q \ln B^q \right) \nonumber \\
&-&
\frac{q^2}{2(q-1)} \,  \sum_{k=1}^n \tr B^{q-1} H^{(k)}(B) \bigg]
\ee
We will apply Lemma \ref{Gross1} to the last term in (\ref{main-pf11}).
As noted before, self-adjointness implies
\bee
\tr B^{q-1} H^{(k)}(B) = \tr B H^{(k)}(B^{q-1})
\eee
Furthermore there is a unitary operator $U_k$ on ${\cal M}_{2^n}$ 
such that
\be\label{main-pf12}
U_k H^{(k)}  U_k^* = {\cal I}_{2^{n-1}} \ot H_k 
\ee
The right side of (\ref{main-pf12}) has the form required for Lemma \ref{Gross1}.
By unitary invariance the same inequality holds for $H^{(k)}$, and thus we
deduce that for each $k=1,\dots,n$
\be\label{main-pf13}
\tr B^{q-1} H^{(k)}(B)
\ge \frac{q-1}{(q/2)^2} \, \tr B^{q/2} H^{(k)}(B^{q/2})
\ee
Applying (\ref{main-pf13}) in
(\ref{main-pf11}) gives the inequality
\be\label{main-pf14}
g'(t) \le \frac{2}{\tr B^q} \, \frac{q-1}{q^2} \, \bigg[ && \hskip-0.35in
n \ln2 \, \tr B^q - \tr B^q \, \ln \tr B^q + \tr B^q \ln B^q \nonumber \\
&-&
2 \, \sum_{k=1}^n \tr B^{q/2} H^{(k)}(B^{q/2}) \bigg]
\ee
and then the log-Sobolev inequality (\ref{main-pf6}) with $A = B^{q/2}$ implies
\bee
g'(t) \le 0
\eee
Thus the left side of (\ref{main-pf10}) is a non-increasing
function of $t$, for all $t \ge 0$. Therefore the inequality (\ref{main-pf10}) holds for all
$t \ge 0$, and this completes the proof.

\par\noindent \qed

{~~}


\begin{thebibliography}{~~}

 \bibitem{Au1} K.M.R. Audenaert,                                                                                                                                                            
``A note on the $p \rightarrow q$ norms of completely positive maps'',
{\em Lin. Alg. Appl} {\bf 430}, 1436 --1440 (2009).

\bibitem{Be} W. Beckner, 
``Inequalities in Fourier analysis'', 
{\em Ann. of Math.} {\bf 102} 
no. 1, 159--182 (1975).

\bibitem{BRW}
A. Ben-Aroya, O. Regev and R. de Wolf,
``A Hypercontractive Inequality for Matrix-Valued Functions
with Applications to Quantum Computing and LDCs'',
{\em Proceedings of the 2008 49th Annual IEEE Symposium on Foundations of Computer Science},
pp. 477--486, 2008.

\bibitem{BFS}
C. H. Bennett, C. A. Fuchs, and J. A. Smolin, 
``Entanglement-Enhanced
Classical Communication on a Noisy Quantum Channel'', 
in {\em Quantum
Communication, Computing and Measurement}, edited by O. Hirota, A. S.
Holevo, and C. M. Caves (Plenum Press, NY, 1997), pages 79Ð88.

 \bibitem{Bi} P. Biane,                                                                                                                                                            
``Free Hypercontractivity'',
{\em Commun. Math. Phys.} {\bf 184}, 457--474, 1997.

\bibitem{Bon} A. Bonami, 
``Etude des coefficients de Fourier des fonctions de $L_p(G)$'',
{\em Ann. Inst. Fourier} {\bf 20}, no. 2 335--402, 1970.

\bibitem{Car1}
E. Carlen,
``Some integral identities and inequalities for entire functions and their application to the
coherent state transform'', {\em Jour. Funct. Anal.}, {\bf 97}, 231--249, 1991.


\bibitem{CL1} E. A. Carlen and E. H. Lieb,
``Optimal hypercontractivity for Fermi fields and related non-commutative integration inequalities'',
{\em Commun. Math. Phys.} {\bf 155}, 27--46, 1993.


\bibitem{DJKR} I. Devetak, M. Junge, C. King and M. B. Ruskai,
``Multiplicativity of completely bounded $p$-norms implies a new additivity result'',
{\em Commun. Math. Phys.} {\bf 266}, 37--63, 2006.

\bibitem{Federbush}
P. Federbush,
``A parially alternate derivation of a result of Nelson'',
{\em Jour. Math. Phys.} {\bf 10}, 50--52, 1969.

\bibitem{Glimm} J. Glimm,
``Boson fields with nonlinear self-interaction in two dimensions'',
{\em Commun. Math. Phys.} {\bf 8}, 12--25 , 1968.

\bibitem{Gr1} L. Gross,
``Logarithmic Sobolev Inequalities'',
{\em American Journal of Mathematics} {\bf 97}, no. 4, 1061--1083, 1975.

\bibitem{Gr2} L. Gross,
``Hypercontractivity and logarithmic Sobolev inequalities for the Clifford-Dirichlet form'',
{\em Duke Math. Jour.} {\bf 43}, 383--396, 1975.

\bibitem{Gr3}
L. Gross,
``Hypercontractivity, Logarithmic Sobolev Inequalities and Applications: A survey of
Surveys'', in
{\em Diffusion, Quantum Theory, and Radically Elementary Mathematics}, ed. W. G. Faris, Princeton University Press, Princeton, NJ, 2006.

\bibitem{HKS}
R. Hoegh-Krohn and B. Simon,
``Hypercontractive semigroups and two dimensional self-coupled Bose fields'',
{\em  Jour. Funct. Anal.}, {\bf 9}, 121--180, 1972.

\bibitem{KKL}
J. Kahn, G. Kalai, and N. Linial,
``The influence of variables on Boolean functions'',
in {\em  Proc. $29$th Annual Symp. Foundations of Computer Science}, pp. 68--80, 1988.

\bibitem{KT}
M. J. Kastoryano and K. Temme,
``Quantum logarithmic Sobolev inequalities and rapid mixing'',
preprint arXiv:1207.3261



 
\bibitem{KR1}
 C. King and M.B. Ruskai,
 ``Minimal Entropy of States Emerging from Noisy Quantum
Channels'',
 {\em IEEE Trans. Info. Theory} {\bf 47}, 1 --19, 2001.
 
 \bibitem{Ki2}
 C. King,
``Inequalities for  trace norms of $2 \times 2$ block matrices'',
{\em Communications in Mathematical Physics} {\bf 242}, 531--545, 2003.
 
 \bibitem{Ki3}
C. King,
``Additivity for unital qubit channels'',
{\em Journal of
Mathematical Physics} {\bf 43}, no. 10, 4641 -- 4653, 2002.


\bibitem{LiebTh} E. H. Lieb and W. Thirring,
``Inequalities for the Moments of the
Eigenvalues of the Schr\"odinger Hamiltonian and Their Relation to Sobolev
Inequalities'', in {\it Studies in Mathematical Physics},  E. Lieb, B.
Simon, A. Wightman eds., pp. 269--303, Princeton University Press, 1976.

\bibitem{M1} A. Montanaro,
``Some applications of hypercontractive inequalities in quantum
information theory'', preprint  	arXiv:1208.0161

\bibitem{MO1}
A. Montanaro and T. Osborne,
``Quantum boolean functions'',
{\em Chicago Journal of Theoretical
Computer Science}, Article 1, 2010.

\bibitem{Nel1}
E. Nelson,
``A quartic interaction in two dimensions'', in {\em Mathematical
Theory of Elementary Particles} (Dedham, Massachusetts, 1965),
R. Goodman and I. E. Segal, eds., pp. 69--73, MIT Press, Cambridge MA,
1966.

\bibitem{Nel2}
E. Nelson,
``The free Markov field'', {\em J. of Funct. Anal.}, {\bf 12} pp. 211--227, 1973.


\bibitem{RSW}
M.B. Ruskai, S. Szarek and W. Werner,
``An Analysis of Completely-Positive Trace-Preserving Maps on $2 \times 2$ Matrices'', 
{\em Lin. Alg. Appl.} {\bf 347}, 159--187, 2002.

\bibitem{Stine} 
W. F. Stinespring, 
``Positive Functions on C*-algebras'', 
{\em Proc. Amer. Math. Soc.} 211--216, 1955.



\bibitem{Wat}
J. Watrous, 
``Notes on super-operator norms induced by Schatten norms'', {\em Quantum
Inf. Comput.} {\bf 5} 57--67, 2005.

\end{thebibliography}
\end{document}